\pdfoutput=1
\documentclass[twocolumn,english,aps,pra]{revtex4-2}

\usepackage[T1]{fontenc}
\usepackage[latin9]{inputenc}
\usepackage{amssymb}
\usepackage{graphicx}
\makeatletter

\providecommand{\tabularnewline}{\\}

\makeatother

\usepackage{babel}

\begin{document}
\title{Phase-space simulations of feedback coherent Ising machines}
\author{Simon Kiesewetter }
\affiliation{Centre for Quantum Science and Technology Theory, Swinburne University
of Technology, Melbourne 3122, Australia}
\author{Peter D Drummond}
\affiliation{Centre for Quantum Science and Technology Theory, Swinburne University
of Technology, Melbourne 3122, Australia}
\begin{abstract}
A new technique is demonstrated for carrying out exact positive-P
phase-space simulations of the coherent Ising machine quantum computer.
By suitable design of the coupling matrix, general hard optimization
problems can be solved. Here, computational quantum simulations of
a feedback type of photonic parametric network are carried out, which
is the implementation of the coherent Ising machine. Results for success
rates are obtained using this scalable phase-space algorithm for quantum
simulations of quantum feedback devices.
\end{abstract}
\maketitle
A wide variety of computational problems can be mapped to the problem
of finding the ground state of an Ising model \citep{lucas2014ising,pakin2017navigating}.
These can be solved with a coherent Ising machine (CIM), a network
of degenerate parametric oscillators (DPOs) operating above their
threshold level, where a single Ising spin is represented by the in-phase
amplitude of the DPO \citep{wang2013coherent,Marandi_CIM_Nature2014,shoji2017quantum,yamamura2017quantum,McMahon_CIM_science2019,Inagaki_CIM_science2019}.
There are two well-established schemes for a DPO-based CIM: one using
coherent optical coupling and one using Field Programmable Gate Array
(FPGA)-based measurement-feedback, as treated here. Each has distinct
advantages and disadvantages \citep{yamamoto2017coherent}. Feedback
methods have led to the development of large photonic quantum computational
devices, with 100,000 nodes being recently demonstrated \citep{honjo2021100}.

A fundamental question is how likely and how fast the CIM will find
its energetic ground state. We wish to analyze this problem from a
fully quantum physical standpoint, where quantum tunneling is potentially
beneficial to the system's performance. For this purpose, it is essential
to develop phase-space simulations, which have been shown to be a
very effective tool for the simulation of large quantum systems and
optical quantum computers  \citep{drummond2016quantum}. Recent examples
of this include up to $16,000$ mode simulations of boson sampling
networks \citep{drummond2021simulating}.

An open quantum system is described by a master equation as a result
of the system Hamiltonian and non-unitary (Lindblad) terms due to
environment interactions. This master equation is converted into a
Fokker-Planck equation (FPE) in phase-space methods (usually a positive-P
or Wigner representation), the solution of which can then be approximated
via sampling its equivalent stochastic differential equation (SDE).

When the system involves feedback that is conditional on measurement,
the conditional master equation contains stochastic terms. These are
due to the quantum noise associated with the monitored observable.
These carry over to the Fokker-Planck equation, resulting in a stochastic
Fokker-Planck equation (SFPE). This cannot immediately be converted
into a set of SDEs, although a weighted approach is possible \citep{hush2009scalable}.
However, in simulating the CIM, one is most interested in the success
probability obtained from the total feedback master equation \citep{wiseman1993quantum,bushev2006feedback},
which does not require weight terms.

It is important to note that the inclusion of quantum feedback may
change the system dynamics profoundly, depending on the specific parameters.
The literature describing systems involving continuous measurements
with simulations via phase-space methods is quite limited, despite
the prominence of feedback control as an experimental technique\citep{zhang2017control}.
Other methods exist for small quantum systems, but the phase-space
techniques are the only ones that have reached large sizes. We have
identified an efficient method in the treatment of quantum feedback,
and applied it to the CIM quantum computer.

An Ising machine is a physical model consisting of a set of effective
spins $\sigma\left(i\right)$, where $\sigma\left(i\right)\in\left\{ -1,1\right\} $
with a set of (real-valued) weights $J_{ij}\in\mathbb{R}$. The system
is designed to reach the ground state of an Ising Hamiltonian: 
\begin{eqnarray}
\mathcal{H}_{IM} & = & -\sum_{i,j}J_{ij}\sigma\left(i\right)\sigma\left(j\right)\,.\label{eq:ising_machine}
\end{eqnarray}

To obtain a nearly equivalent experimental model, a driven and damped
feedback network is used. Each node $j$ is a degenerate optical parametric
oscillator (DOPO) \citep{Drummond_OpActa1981,Sun_NJP2019,Sun_PRA2019,teh2020dynamics},
treated here as an optical cavity that contains two modes $a$ (the
signal mode) and $a_{p}$ (the pump mode), where in our case $\omega_{p}=2\omega_{s}$.
In the rotating reference frame, the Hamiltonian is 
\begin{eqnarray}
\mathcal{H} & = & \mathcal{H}_{int}+\mathcal{H}_{pump}+\mathcal{H}_{loss},
\end{eqnarray}
where $\mathcal{H}_{int},\mathcal{H}_{pump},\mathcal{H}_{loss}$ denote
the interaction, pump and loss terms, respectively and 
\begin{eqnarray}
\mathcal{H}_{int} & = & \frac{i\hbar\kappa}{2}\left[\left(a^{\dagger}\right)^{2}a_{p}-a^{2}a_{p}^{\dagger}\right]\nonumber \\
\mathcal{H}_{pump} & = & i\hbar\left[\epsilon_{p}a_{p}^{\dagger}-\epsilon_{p}^{*}a_{p}\right]
\end{eqnarray}
Here $\epsilon_{p}$ is the pump field amplitude and $\kappa$ denotes
the non-linear interaction \citep{drummond2014quantum}.

As well as the usual pump and loss terms common to equations of this
type, the experiment requires a coupling between the oscillators in
order to implement an Ising-like model, which includes a feedback
loop. This enables the output of one DOPO to be used to modify the
pump rate in the other oscillators. The overall effect of this leads
to an Ising-like coupling between the oscillators. A feedback-controlled
system such as the DOPO network studied here therefore requires a
measurement carried out continuously on its system state. Considering
the state collapse that comes with every measurement on a quantum
system, it is not obvious at all how a continuous measurement, resulting
in a ``continuous state collapse'' can be described mathematically.

This question has been addressed in the quantum optics literature
\citep{Milburn1987quantum,dalibard1992wave,wiseman1993quantum,wiseman1994quantum,jacobs2006straightforward,wiseman2009quantum},
with later extensions \citep{hush2009scalable}. One considers a monitored
quantum system described by a conditional density operator $\rho_{c}$,
that depends on a set of measurement outcomes. A feedback master equation
in the Ito stochastic calculus is obtained. A measurement noise is
generated during a specific \textit{realization} of the measurement.
However, we are more interested in the system evolution averaged over
all possible measurement noise realizations, which is described by
the much simpler total master equation.

For this case, we recursively extend the single homodyne detection
result\citep{wiseman1994quantum} to treat multiple feedback loops,
giving:
\begin{equation}
\dot{\rho}=\mathcal{L}\rho+\sum_{j}\mathcal{K}_{j}\left(a_{j}\rho+\rho a_{j}^{\dagger}\right)+\frac{1}{4\gamma_{m}}\sum_{\mu}\mathcal{K}_{j}^{2}\rho.
\end{equation}
Here $\mathcal{L}\rho$ described unmonitored evolution and total
damping, while the $\mathcal{K}_{j}$ super-operators describe the
feedback of the measured quadrature $X_{j}=a_{j}+a_{j}^{\dagger}$
onto the system, with a corresponding amplitude decay rate of $\gamma_{m}$
including measurement efficiency factors.

This averages the conditional Ito master equation over all the noises,
which is equivalent to setting the noise terms to zero \citep{wiseman1993quantum,bushev2006feedback},
owing to the factorization properties of Ito stochastic equations
combined with the linearity of the stochastic master equation as a
function of $\rho_{c}$.

We take the $j$-th feedback super-operator $\mathcal{K}_{j}$ to
be:
\begin{equation}
\mathcal{K}_{j}\rho=\zeta\sum_{i}J_{ij}\left[a_{i}^{\dagger}-a_{i},\rho\right]\,,
\end{equation}
where $\zeta$ is a feedback factor, and $J_{ij}$ is the Ising model
matrix. Combining this feedback treatment with the theory of each
single degenerate optical parametric oscillator \citep{Drummond_OpActa1981},
the simplest quantum model of the $N$-node CIM is therefore described
by the total master equation
\begin{eqnarray}
\frac{d\rho}{dt} & = & \sum_{i}\left\{ \gamma\left(2a_{i}\rho a_{i}^{\dagger}-a_{i}^{\dagger}a_{i}\rho-\rho a_{i}^{\dagger}a_{i}\right)\right.\nonumber \\
 &  & +\gamma_{p}\left(2a_{pi}\rho a_{pi}^{\dagger}-a_{pi}^{\dagger}a_{pi}\rho-\rho a_{pi}^{\dagger}a_{pi}\right)\nonumber \\
 &  & +\left[\epsilon_{pi}a_{pi}^{\dagger}-h.c,\rho\right]+\frac{1}{4\gamma_{m}}\mathcal{K}_{i}^{2}\rho\nonumber \\
 &  & \left.+\frac{\kappa}{2}\left[\left(a_{i}^{\dagger}\right)^{2}a_{pi}-a_{i}^{2}a_{pi}^{\dagger},\rho\right]+\mathcal{K}_{i}\left(a_{i}\rho+\rho a_{i}^{\dagger}\right)\right\} \,.\label{eq:mast_eq}
\end{eqnarray}

The signal and pump modes $a_{i}$ and $a_{pi}$ are associated with
unmonitored decay rates $\gamma_{s},\gamma_{p}$, respectively. There
is an additional system parameter $\gamma_{m}$, representing measurement
of the signal mode, while $\gamma$ is the total decay rate including
measurement loss, and $\mathcal{K}_{j}$ is the feedback super-operator
given above. The second-order term in $\mathcal{K}_{j}$ in (\ref{eq:mast_eq})
accounts for the quantum diffusion introduced by the injection of
quantum noise during the feedback process, and is an important source
of decoherence.

The pump strength $\epsilon_{pi}$ is independent of the signal amplitude
and will generally be set as uniform for all DOPOs ($\epsilon_{pi}=\epsilon_{p}$).
The action of the feedback term in Eq. (\ref{eq:mast_eq}) is similar
to inducing a signal input $\hat{\epsilon}_{i}$ which results from
continuous homodyne detection via 
\begin{eqnarray}
\hat{\epsilon}_{i} & = & \zeta\sum_{j}J_{ij}\hat{X}_{j}\,,\label{eq:Esi}
\end{eqnarray}
where $\hat{X}_{j}=a_{j}^{\dagger}+a_{j}$ is the measured quadrature
for DOPO site $j$. It is also subject to measurement quantum noise,
but this averages to zero for the total master equation, and is therefore
omitted in such equations.

While master equations can be treated directly using a number state
expansion, the exponentially large size of the Hilbert space makes
this computationally prohibitive for large network experiments. This
necessitates alternative approaches to simulation. 

One approach is to include measurement noise terms, giving a conditional
master equation. This does not translate into a standard FPE, but
to a stochastic Fokker-Planck equation (SFPE), which cannot be converted
into a conventional stochastic differential equation (SDE). The noise
terms associated with the measurement are different in nature to the
familiar noise terms one typically encounters for the integration
of SDEs. In the literature, these measurement noise terms are often
called ``physical'' noise. The approach used in some earlier papers
is to use weighted phase-space trajectories to treat them. While this
allows a detailed understanding of the measurement noise, it reduces
the simulation efficiency.

Here we describe an exact phase-space method that maps the \emph{total
}quantum master equation into intuitively understandable and readily
simulated stochastic equations. This ignores the conditional current,
but it reliably gives the average final state of the system. Therefore,
Eq. (\ref{eq:mast_eq}) can be very efficiently treated using phase-space
methods. The total feedback master equation requires no additional
weighting terms, and is especially suitable for treating very large
quantum feedback systems, due to its greater scalability. Taking an
average over the measurement noise does not change the overall master
equation or the success rate of the CIM.

The positive-P phase-space representation \citep{Drummond_generalizedP1980}
gives an exact method for integrating such master equations provided
boundary terms vanish. This requires that the probabilities decay
rapidly enough at large radius in phase-space \citep{Gilchrist_PRA1997},
which is well-satisfied here. Another method, the truncated Wigner
method\citep{wigner1932quantum,drummond1993simulation}, truncates
terms of O($N^{-3/2})$ in an expansion in the photon number $N$,
to obtain a positive-valued distribution. Due to the small number
of photons at the initial stages of the experiment, we use the more
accurate positive-P approach instead.

Eq. (\ref{eq:mast_eq}) results in the Fokker-Planck equation (FPE)
for the positive-P representation: 
\begin{equation}
\frac{dP}{dt}=\sum_{i}\left[\mathcal{D}_{i}-2\zeta\sum_{j}J_{ij}\partial_{x_{i}}\left[x_{j}-\frac{\zeta}{2\gamma_{m}}\sum_{k}J_{kj}\partial_{x_{k}}\right]\right]P\,.\label{eq:posP_SFPE-1}
\end{equation}

Here, $\mathcal{D}_{i}$ is the standard OPO Fokker-Planck equation
for a single device, with the damping rate given by $\gamma$, and
we define $x_{j}\equiv\alpha_{j}+\beta_{j}$ and $\partial_{x_{i}}\equiv\left(\partial_{\alpha_{i}}+\partial_{\beta_{i}}\right)/2$.
The resulting Ito stochastic differential equations are: 
\begin{eqnarray}
\dot{\alpha}_{i} & = & \epsilon_{i}-\gamma\alpha_{i}+\kappa\beta_{i}\alpha_{pi}+\sqrt{\kappa\alpha_{pi}}\xi_{i}^{\alpha}\nonumber \\
\dot{\beta}_{i} & = & \epsilon_{i}-\gamma\beta_{i}+\kappa\alpha_{i}\beta_{pi}+\sqrt{\kappa\beta_{pi}}\xi_{i}^{\beta}\nonumber \\
\dot{\alpha}_{pi} & = & \epsilon_{p}-\gamma_{p}\alpha_{pi}-\frac{\kappa}{2}\alpha_{i}^{2}\nonumber \\
\dot{\beta}_{pi} & = & \epsilon_{p}-\gamma_{p}\beta_{pi}-\frac{\kappa}{2}\beta_{i}^{2}\,.\label{eq:SDE_posP-1}
\end{eqnarray}
The stochastic feedback term $\epsilon_{i}$ includes both a coherent
term and a noise term coming from the second order derivative terms
in the Fokker-Planck equation, so that: 
\begin{equation}
\epsilon_{i}\equiv\zeta\sum_{j}J_{ij}\left(\alpha_{j}+\beta_{j}+\frac{1}{\sqrt{2\gamma_{m}}}\xi_{j}\right),
\end{equation}
where $\xi_{i}^{\alpha}$, $\xi_{i}^{\beta}$, $\xi_{i}$ are delta-correlated
Gaussian noise increments.

Next, we look at the adiabatically eliminated SDEs in Eq. (\ref{eq:SDE_posP-1}),
obtained by setting $\dot{\alpha}_{pi}=\dot{\beta}_{pi}=0$ in the
Ito stochastic equations \citep{Drummond_OpActa1981,gardiner1984adiabatic}.
Identical results are obtained, although with greater complexity,
using operator adiabatic methods within the master equation \citep{carmichael2009statistical}.
We introduce an effective nonlinearity, $\chi\left(\alpha\right)\equiv\kappa\left[\epsilon_{p}-\frac{\kappa}{2}\alpha^{2}\right]/\gamma_{p}$
, and utilize this to find the stochastic differential equations.
For greater efficiency in numerical integration \citep{drummond1991computer},
these can be transformed into their equivalent Stratonovich form.
This modifies the damping rate so that $\gamma\rightarrow\gamma'\equiv\gamma-\kappa^{2}/4\gamma_{p}\,$,
giving: 
\begin{eqnarray}
\dot{\alpha}_{i} & = & \epsilon_{i}-\gamma'\alpha_{i}+\beta_{i}\chi\left(\alpha_{i}\right)+\sqrt{\chi\left(\alpha_{i}\right)}\xi_{i}^{\alpha}\nonumber \\
\dot{\beta}_{i} & = & \epsilon_{i}-\gamma'\beta_{i}+\alpha_{i}\chi\left(\beta_{i}\right)+\sqrt{\chi\left(\beta_{i}\right)}\xi_{i}^{\beta}\,.\label{eq:SDE_Strat}
\end{eqnarray}
The stochastic nature of the feedback noise is included in the feedback
term $\epsilon_{i}$, while $\xi_{i}^{\alpha,\beta}$ are internal
quantum noise terms that generate non-classical squeezing and entanglement.

We first consider an experiment with 2 coupled DOPOs. Here, the interaction
matrix $\mathbf{J}$ is simply $J_{ij}=\left(\delta_{ij}-1\right)$,
corresponding to antiferromagnetic coupling. The experiment is initialized
with both cavity modes $a_{1,2}$ in the vacuum state. During the
simulation time, which extends from $0$ to $T_{max}=100$, the pump
strength $\epsilon_{p}$ is linearly increased from $0$ to $2\cdot\epsilon_{th}$,
where $\epsilon_{th}=\gamma\gamma_{p}/\kappa$ is the threshold pump
strength. This way, the system is gradually steered from a below-threshold
to an above-threshold region where different spin configurations correspond
to local minima in the energy landscape. A gradual transition increases
the likelihood of finding a global minimum corresponding to the two
degenerate ground states of the Ising system simulated here, closely
resembling a simulated annealing process. The system parameters are
$\gamma=1.1$, $\gamma_{p}=100$, $\kappa=0.316227$, $\gamma_{m}=0.1$,
$\zeta=0.3$. For the positive-P integration, we used a hierarchy
of $N_{e}\times N_{s}$ ensembles, in order to calculate the sampling
errors \citep{kiesewetter2016xspde}, with $N_{s}=8192$ subensemble
samples and $N_{t}=50\cdot10^{3}$ time-steps. We integrated the system
dynamics according to Eqs. (\ref{eq:SDE_Strat}) using a stochastic
RK4 scheme. The simulations were run on a computer cluster using multiple
GPUs and implemented in C++ using CUDA. An independent check was carried
out using a public domain stochastic integration code \citep{kiesewetter2016xspde,opanchuk2018simulating},
with identical results. Discretization error and sampling errors were
checked, and found to be negligible. We define a success rate as the
number of quantum trajectories that adopt one of the two degenerate
ground states of the antiferromagnetic Ising model divided by the
total number of quantum trajectories. The ground states are indicated
by the sign signatures $\left(+,-\right)$ and $\left(-,+\right)$
for the $x$-quadratures of the cavity modes $a_{1,2}$. A success
rate of $1$ is obtained for $100\times8192$ independent repetitions
of the simulation.

\begin{figure}
\begin{tabular}{|c|}
\hline 
\includegraphics[width=0.4\textwidth]{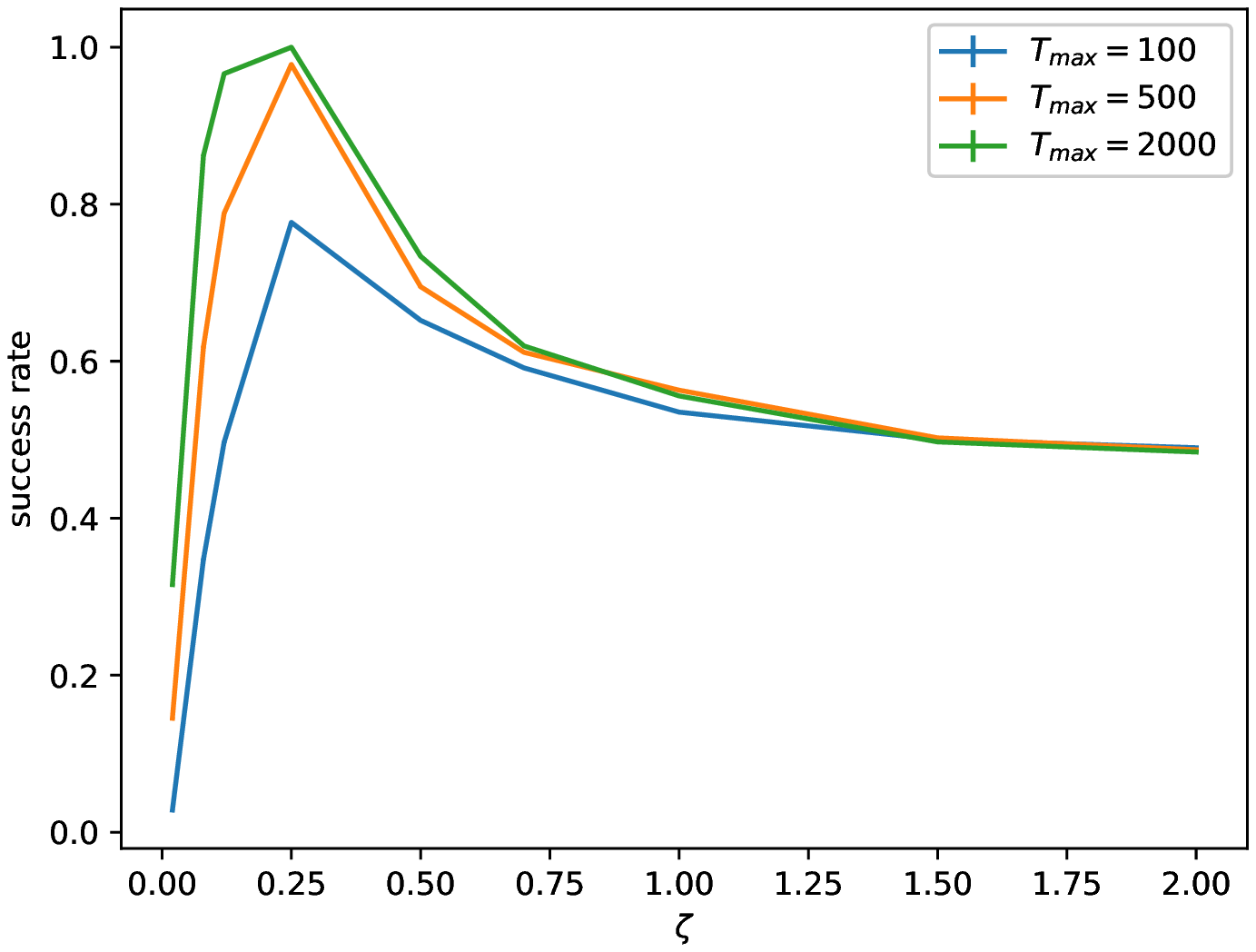}\tabularnewline
\hline 
\end{tabular}\caption{\label{fig:N16DOPO}Simulation results for the 16-DOPO experiment,
showing the success rate for two different integration times and a
range of interaction parameters $\zeta$. The results were obtained
using the positive-P scheme, using $20\times8192$ independent simulations
at each data point for $T_{max}=100$, $T_{max}=500$ and $T_{max}=2000$,
respectively.}
\label{Fig1}
\end{figure}

\begin{figure}
\begin{tabular}{|c|}
\hline 
\includegraphics[width=0.4\textwidth]{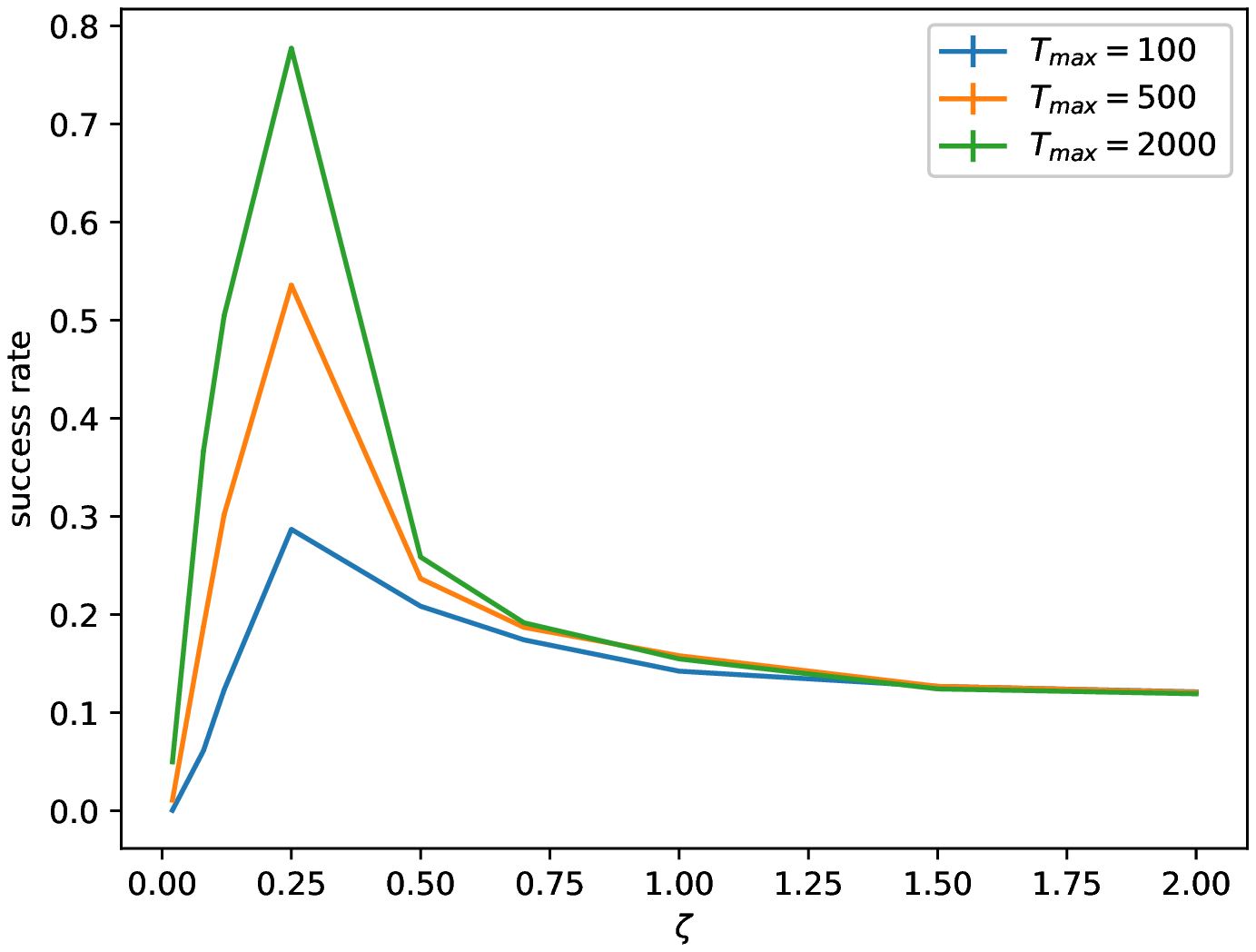}\tabularnewline
\hline 
\end{tabular}\caption{\label{fig:N32DOPO}Simulation results for the 32-DOPO experiment,
showing the success rate for two different integration times and a
range of interaction parameters$\zeta$, using $20\times8192$ independent
simulations at each data point.}
\label{Fig2}
\end{figure}

We also considered two larger experiments, consisting of $N=16$ and
$N=32$ DPO cavities, respectively, which are more realistic cases.
The cavities are coupled antiferromagnetically through nearest neighbor
interaction in a circular sense where $\mathbf{J}$ is given by $J_{ij}=-1$
if $\left|i-j\right|=1$ or $\left|i-j\right|=N-1$, $J_{ij}=0$ otherwise.
The ground states have sign signatures $\left(+,-,...,+,-\right)$
and $\left(-,+,...,-,+\right)$ for the $x$-quadratures of $a_{1},...,a_{N}$.

The parameters $\gamma$, $\gamma_{p}$, $\kappa$, $\gamma_{m}$,
$N_{S}$, $N_{t}$ are the same as in the $N=2$ case, however, the
interaction strength $\zeta$ is varied. We also use three different
integration times $T_{max}$ to demonstrate its effect on the simulation
efficacy of the simulated coherent Ising machine. We integrate the
system dynamics and determine the success rate analogously to the
$N=2$ case, which is plotted against the interaction strength $\zeta$
for a fixed integration time $T_{max}$. We have simulated $9$ different
values for $\zeta$. The results are shown in Figs. \ref{Fig1} and
\ref{Fig2}. Generating all data points for $T_{max}$ and $\zeta$
takes about 1 hour, when utilizing 10 GPUs (NVIDIA P100).

From these results, very high success rates close to $1$ can be achieved
with the right choice of $T_{max}$ and $\zeta$, although this requires
further investigation into different types of optimization problem.
Longer simulation times, with an adiabatic increase of the pump field,
result in a higher success rate. These simulations also allow other
types of sweep to be investigated. Somewhat counter-intuitively, a
strong feedback coupling does not always improve the results, and
in may reduce the success rate. This appears to be caused by the quantum
feedback noise introducing errors due to random jumps between the
near ground-state solutions, resulting in an effective thermalization
of the steady-state. Between the $N=16$  and $N=32$ case, the success
rate is reduced significantly, especially for non-optimal values of
$\zeta$. A decrease of the success rate with increasing problem size
is consistent with previous findings for similar CIM systems.

In conclusion, we have derived and implemented a numerical scheme
for accurately simulating the coherent Ising machine in a feedback
implementation. The innovation terms are treated in a fully mathematically
justified way, proven to yield the exact expectation values in the
limit of a large number of independent simulations. This will be an
even more important feature when dealing with increasingly nonclassical
cases. Since the methods are exact, they can be used as benchmarks
for developing faster or more accurate algorithms in future. We note
that a large variety of parameter values and sweep types are possible.
These will be the subject of subsequent investigations.

\section*{\textup{\normalsize{}{}Acknowledgements}}

This work was performed on the OzSTAR national facility at Swinburne
University of Technology. OzSTAR is funded by Swinburne University
of Technology and the National Collaborative Research Infrastructure
Strategy (NCRIS). This work was funded through a grant from NTT Phi
Laboratories.

\bibliographystyle{apsrev4-2}
\bibliography{CIM}

\end{document}